\documentclass[11pt]{article}

\usepackage[utf8]{inputenc}
\usepackage[T1]{fontenc}
\usepackage{lmodern}
\usepackage{microtype}
\usepackage{geometry}
\usepackage{graphicx}
\usepackage{booktabs}
\usepackage{array}
\usepackage{enumitem}
\usepackage{xcolor}
\usepackage{hyperref}
\usepackage{tikz}
\usepackage{orcidlink}
\usetikzlibrary{arrows.meta,positioning,fit,shapes.multipart}
\usepackage[round,authoryear]{natbib}

\hypersetup{
  colorlinks=true,
  linkcolor=blue!50!black,
  citecolor=blue!50!black,
  urlcolor=blue!50!black,
  pdftitle={Towards an Interactive Evidence-RAG Peer-Review Workspace for the Journal of Digital History},
  pdfauthor={Elisabeth Guerard, Mehrdad Almasi, Marion Salaun, Frederic Clavert,Mirjam Pfeiffer}
}

\title{Towards an Interactive Evidence-RAG Peer-Review Workspace for the Journal of Digital History\\
\large Preliminary Version of a Full Paper}

\author{
Élisabeth Guérard\,\orcidlink{0000-0001-7742-4141},
Mehrdad Almasi\,\orcidlink{0000-0001-8203-0606},
Marion Salaün\,\orcidlink{0009-0005-3624-823X},
Frédéric Clavert\,\orcidlink{0000-0002-0237-2532},
Mirjam Pfeiffer\,\orcidlink{0009-0005-2265-5926}\\
Luxembourg Centre for Contemporary and Digital History (C\textsuperscript{2}DH), University of Luxembourg
}

\begin{document}
\maketitle

\begin{abstract}
This preliminary paper presents an interactive \emph{Evidence-RAG} workspace for editorial assessment of AI-assisted peer review in the \emph{Journal of Digital History}. The workflow makes model recommendations easier to inspect by linking reviewer comments, paper evidence, retrieval traces, and reproducibility checks. The system does not replace editors or reviewers. It treats large language models as auditable assistants whose outputs must be checked by human scholars. We describe the current pipeline: paper conversion, semantic chunking, vector indexing, retrieval-augmented evidence assessment, and a lightweight editorial interface. This is a preliminary version of a full paper associated with the accepted presentation ``Towards an Interactive `Evidence-RAG' Peer-Review Workspace for the Journal of Digital History'' at the conference \emph{AI through History, History through AI}, C\textsuperscript{2}DH, University of Luxembourg, 15--16 June 2026. The work was submitted to the conference on 26 February 2026. We also report a first editor-annotation analysis of 80 saved decisions for the Claude-Qwen audit configuration. For Claude-Qwen, strict editor-confirmed accuracy is 70.0\%, the correct-or-mostly-correct rate is 86.2\%. The useful-output rate includes all responses judged correct, mostly correct, or partially correct, since these outputs can still assist with editorial review even if the model’s assessment is not fully accurate (this rate was 90.0\%). The full version may include additional experiments, extended evaluation, and a more complete release of materials.
\end{abstract}


\section{Introduction}
Peer review is central to scholarly communication. However, editorial decisions are often made under time pressure, with uneven reviewer coverage and limited traceability between comments and manuscript evidence. Large language models (LLMs) can summarise, classify, and compare scholarly texts, but their recommendations are not always easy to audit. This is a serious issue in editorial work, because a confident model judgement may still be based on weak retrieval, mismatched evidence, or unsupported claims.

This paper presents a preliminary version of an \emph{Evidence-RAG} workspace developed around the \emph{Journal of Digital History} (JDH). The aim is not to automate editorial decisions. The aim is to make AI-assisted peer-review support more transparent by linking reviewer comments to evidence retrieved from the paper, showing the model's support judgement, and keeping the editor as the final decision-maker.

The contribution of this preliminary paper is threefold. First, it describes a modular workflow that turns submissions and reviews into auditable evidence objects. Second, it proposes an editorial interface where model labels and confidence components can be inspected rather than accepted automatically. Third, it treats reproducibility as an editorial requirement: each model recommendation should be connected to retrieved chunks, explicit scoring components, and metadata that can be reviewed later.

\section{Status of this Version}
This manuscript is intentionally concise. It is a preliminary arXiv version prepared before the complete article. The associated presentation has been accepted in the programme of \emph{AI through History, History through AI}, under Panel V, ``Infrastructure, Workflow \& Hermeneutic AI,'' on Tuesday, 16 June 2026. The work was submitted to the conference on 26 February 2026. The author list, experimental scope, and evaluation tables may be extended in a later full version.

\section{Related Work and Motivation}
The \emph{Journal of Digital History} combines narrative argument, hermeneutic interpretation, and code-and-data layers. This publication model is rich, but it also makes peer review more difficult. A single reviewer may not be able to evaluate interpretive claims, computational methods, and reproducibility materials with the same level of confidence. This motivates a workspace that supports human editorial judgement without replacing it.

AI assistance is already becoming part of peer-review workflows. Survey data indicates that many researchers rely on AI during peer review, at times without acknowledging its use or even in conflict with stated policies \citep{naddaf2025peerreview}. Journals and conferences are similarly testing AI as both authors and reviewers \citep{bianchi2026agents4science}. LLM-based reviewing agents can deliver fast, well-structured comments to support authors or human reviewers \citep{renata2025ai,bougie2025generative}. Multi-agent and multi-role frameworks likewise attempt to emulate aspects of peer review and benchmark AI-generated reports against human evaluations \citep{gao2025reviewagents,bharti2026coreviewer,rajakumar2026peerreview, zhan2026let}. However, the fluency of AI-produced text can obscure underlying flaws. For instance, studies of AI-generated code contributions show that reviewers may react neutrally or even favorably despite unresolved issues of quality and reuse \citep{huang2026code}. These risks make auditability crucial.

This research addresses these challenges through an evidence-bounded RAG workflow. Rather than requesting that a model assess a full manuscript in a non-transparent way, the system links each reviewer comment to specific retrieved passages, records both the model's judgement and its confidence, and presents these elements to editors as inspectable evidence objects. This design aligns with claim--evidence interfaces in scholarly question answering, which anchor model outputs in visible provenance instead of untraceable model generation \citep{martinboyle2026papertrail}. The guiding governance principle is that AI may assist in peer review only if prompts, model versions, confidence scores, and any departures from checklist-based assessments remain fully transparent and accountable \citep{sen2025rigor}.
\section{Workflow Overview}
The current workflow is organised as a sequence of reproducible stages. A central design choice is that the model does not receive the full paper as one context. Instead, the system retrieves limited evidence chunks and asks the model to assess a reviewer comment against those passages. This RAG-based design reduces unnecessary exposure of manuscript content. It also creates room for future experiments with commercial models, if privacy, contractual, and institutional safeguards allow it. In the implementation, file names begin with a digit to indicate the order of the stages. The project also includes two stage-three notebooks/scripts because two different model configurations and execution settings were tested.

\begin{figure}[t]
\centering
\includegraphics[width=\linewidth]{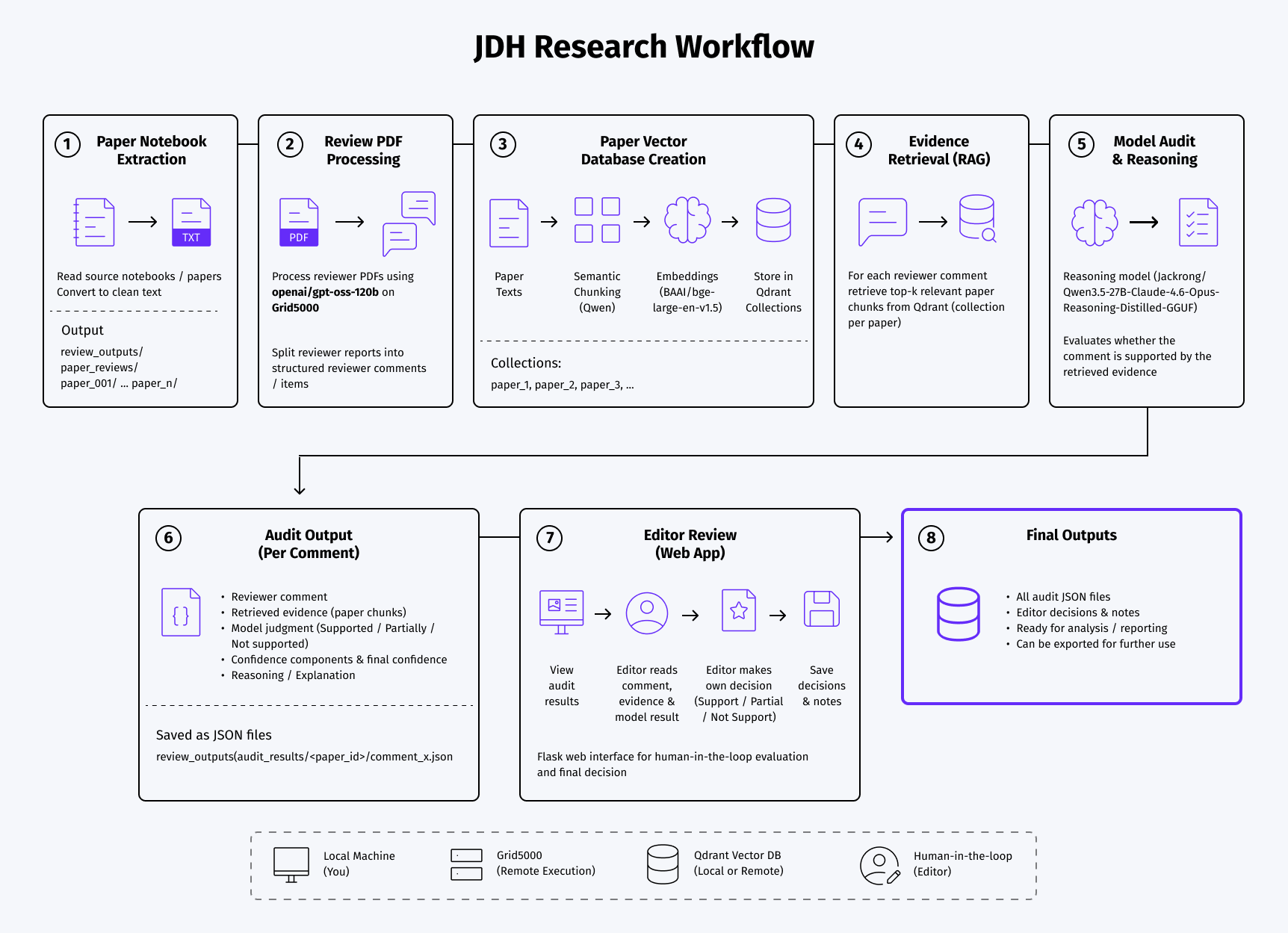}
\caption{JDH research workflow for the Evidence-RAG peer-review workspace.}
\label{fig:jdh-research-workflow}
\end{figure}

\paragraph{Stage 0: Paper Conversion.}
The first stage converts papers into machine-readable text. The goal is to keep enough section information for later evidence mapping while avoiding unnecessary changes to the manuscript.

\paragraph{Stage 1: Review Processing.}
Reviewer comments are separated and normalised. Each comment is then treated as an individual claim or judgement that requires evidence. This stage prepares the unit of analysis for retrieval and model assessment.

\paragraph{Stage 2: Vector Database Construction.}
The paper text is divided into sections and semantically coherent chunks. The implementation uses an embedding model and a vector database so that each reviewer comment can be matched with candidate evidence passages. The current implementation uses BAAI/bge-large-en-v1.5 embeddings with 1024 dimensions and Qdrant as the vector store.

\paragraph{Stage 3: Evidence-RAG Auditing.}
The Evidence-RAG stage retrieves paper chunks for each reviewer comment. It then asks a model to judge whether the retrieved evidence supports the comment. The prompt is evidence-bounded: it contains the reviewer comment, selected retrieved passages, and structured instructions, not the complete manuscript. The output is an audit object with the model label, retrieved evidence, scoring components, and metadata. In the current prototype, labels include \texttt{supported}, \texttt{partially\_supported}, \texttt{not\_supported}, and \texttt{insufficient\_evidence}.

\section{Editorial Workspace}
The editor-facing workspace is designed for inspection, not automation. Users can filter by paper and model, read the reviewer comment, inspect the model recommendation, and compare the recommendation with retrieved evidence. The interface stores the editor's decision separately from the model output, so human judgement remains the final authority.

\begin{table}[t]
\centering
\small
\begin{tabular}{p{0.26\linewidth}p{0.64\linewidth}}
\toprule
\textbf{Element} & \textbf{Purpose} \\
\midrule
Reviewer comment & Unit of editorial assessment. \\
Retrieved chunks & Paper passages used as evidence for or against the comment. \\
Model label & Model judgement about whether the comment is supported by the evidence. \\
Confidence components & Interpretable indicators of retrieval strength, model support, evidence specificity, and section coverage. \\
Editor decision & Final human judgement, stored independently from the model label. \\
\bottomrule
\end{tabular}
\caption{Core objects displayed in the preliminary editorial workspace.}
\label{tab:workspace}
\end{table}

\section{Confidence Components}
The current prototype uses a transparent composite confidence score. This score is not a universal measure of truth. It is a practical signal that helps editors decide where to focus attention:
\begin{equation}
\mathrm{Confidence} = 0.15R + 0.25M + 0.45E + 0.15S,
\end{equation}
where $R$ is retrieval strength, $M$ is model support strength, $E$ is evidence specificity, and $S$ is section coverage. Each component is normalised to a value between 0 and 1.

\begin{itemize}
    \item \textbf{Retrieval strength ($R$)} measures how strongly the retrieved chunks match the reviewer comment according to the retrieval scores. It is based on the average similarity score of the retrieved evidence passages.
    \item \textbf{Model support strength ($M$)} represents how strongly the model judges that the reviewer comment is supported by the retrieved evidence. Higher values indicate stronger model confidence in the support label.
    \item \textbf{Evidence specificity ($E$)} measures how directly and specifically the retrieved passages address the reviewer comment. This component captures whether the evidence is not only semantically close, but also substantively relevant.
    \item \textbf{Section coverage ($S$)} measures how broadly the retrieved evidence is distributed across the paper. It is calculated from the number of unique retrieved sections relative to the total number of sections in the paper.
\end{itemize}

The largest weight is given to evidence specificity because the project focuses on whether the retrieved passages are relevant to the reviewer comment. In later versions, these weights should be tested empirically and may be revised.
\section{Infrastructure and Reproducibility}
The prototype was designed to run on controlled research infrastructure, including Grid'5000 \cite{grid5000team2026} for model processing. This is important for two reasons. First, submissions and reviews may contain sensitive editorial data. Second, reproducibility requires stable job scripts, explicit model identifiers, and recorded outputs. Because the system is RAG-based, future deployments could compare local open models with commercial models without sending entire papers, if only necessary retrieved chunks are transmitted and governance rules permit it. The current implementation records model names, embedding settings, chunk manifests, vector collections, audit files, and job summaries.

For the preliminary workflow, Grid'5000 provides a private and reliable European research infrastructure for controlled experimentation. The pipeline stores intermediate outputs under a project-local directory structure, downloads models to a persistent model store, and records job status information to reduce silent failures.

\section{Preliminary Evaluation and Analysis}
This version includes an editor-annotation export from the workspace  containing 80 saved editor decisions across three papers, all from the Claude-Qwen audit configuration. We report three rates. \emph{Strict accuracy} counts only editor decisions marked \emph{correct}. The \emph{correct-or-mostly-correct rate} counts \emph{correct} and \emph{mostly correct}. The broader \emph{useful-output rate} also includes \emph{partly correct} cases, because these outputs may still help editors locate an issue even when the recommendation is incomplete. All 80 decisions in the corrected export belong to the Claude-Qwen audit configuration.

\begin{table}[t]
\centering
\small
\begin{tabular}{lrrrrrrr}
\toprule
\textbf{Subset} & \textbf{Items} & \textbf{Correct} & \textbf{Mostly} & \textbf{Partly} & \textbf{Incorrect} & \textbf{Strict acc.} & \textbf{Useful rate} \\
\midrule
Claude-Qwen editor decisions & 80 & 56 & 13 & 3 & 8 & 70.0\% & 90.0\% \\
\bottomrule
\end{tabular}
\caption{Preliminary editor-decision counts and accuracy rates from the corrected evaluation export. The useful rate is calculated as correct + mostly correct + partly correct divided by all items.}
\label{tab:model-summary}
\end{table}

\begin{table}[t]
\centering
\small
\begin{tabular}{lrrrrrrr}
\toprule
\textbf{Claude-Qwen subset} & \textbf{Items} & \textbf{Correct} & \textbf{Mostly} & \textbf{Partly} & \textbf{Incorrect} & \textbf{Strict acc.} & \textbf{Correct+mostly} \\
\midrule
All Claude-Qwen papers & 80 & 56 & 13 & 3 & 8 & 70.0\% & 86.2\% \\
Paper 1 & 32 & 20 & 6 & 2 & 4 & 62.5\% & 81.2\% \\
Paper 2 & 16 & 13 & 3 & 0 & 0 & 81.2\% & 100.0\% \\
Paper 3 & 32 & 23 & 4 & 1 & 4 & 71.9\% & 84.4\% \\
\bottomrule
\end{tabular}
\caption{Paper-level preliminary evaluation for the Claude-Qwen Evidence-RAG audit configuration. Strict accuracy is correct divided by all items. Correct+mostly counts correct and mostly correct decisions.}
\label{tab:claude-qwen-by-paper}
\end{table}

For Claude-Qwen, 56 of 80 decisions were marked \emph{correct}, giving a strict accuracy of 70.0\%. Another 13 decisions were marked \emph{mostly correct}. Together, 69 of 80 decisions are correct or mostly correct, which is 86.2\%. If the three \emph{partly correct} cases are also counted as useful but incomplete outputs, the result is 72 of 80, or 90.0\%.

At the paper level, Paper 2 has the strongest result: all 16 decisions were either correct or mostly correct. Paper 1 includes 20 correct, six mostly correct, two partly correct, and four incorrect decisions. This gives 62.5\% strict accuracy, 81.2\% correct-or-mostly-correct, and 87.5\% useful-output rate. Paper 3 includes 23 correct, four mostly correct, one partly correct, and four incorrect decisions. This gives 71.9\% strict accuracy, 84.4\% correct-or-mostly-correct, and 87.5\% useful-output rate.

The editor notes show both the value and the limits of the evidence-bounded RAG approach. Some necessary evidence was not retrieved. Some reviewer comments combined questions, claims, and suggestions that should be split into smaller units. In addition, many reviewer suggestions cannot be verified directly, because Evidence-RAG checks what is present in the submitted work rather than deciding whether a proposed improvement is desirable. These observations support the central design principle of the workspace: model output should guide inspection, but the editor must make the final judgement.

The full paper will extend this analysis in four directions: a larger annotated corpus; systematic comparison between open and commercial model configurations; retrieval diagnostics by paper section and comment type; and qualitative failure analysis of unsupported comments, missing evidence, and overconfident rationales.

\section{Additional Audit-File Statistics}
This additional analysis uses the three Claude-Qwen audit JSON files directly. It is separate from the editor-accuracy analysis above. The aim is to describe the audit outputs themselves: model support labels, confidence components, and retrieval behaviour. Across the three audit files, there are 80 comment-level audit records. Paper 1 contains 32 records, Paper 2 contains 16 records, and Paper 3 contains 32 records. Each audit record contains five retrieved chunks, giving 400 retrieved chunks in total. The most frequent retrieved section was \texttt{[CODE]} with 199 retrieved chunks, followed by Discussion with 55, Conclusion with 48, Introduction with 35, and Results with 28. The remaining 35 retrieved chunks came from other sections or section labels.

\begin{table}[t]
\centering
\small
\begin{tabular}{lrrrrrrrr}
\toprule
\textbf{Subset} & \textbf{Items} & \textbf{Sup.} & \textbf{Part.} & \textbf{Not} & \textbf{Insuf.} & \textbf{Conf.} & \textbf{Retr.} & \textbf{Spec.} \\
\midrule
Paper 1 & 32 & 8 & 12 & 11 & 1 & 0.630 & 0.696 & 0.536 \\
Paper 2 & 16 & 2 & 9 & 5 & 0 & 0.655 & 0.724 & 0.620 \\
Paper 3 & 32 & 3 & 11 & 17 & 1 & 0.569 & 0.649 & 0.429 \\
\midrule
All papers & 80 & 13 & 32 & 33 & 2 & 0.611 & 0.683 & 0.510 \\
\bottomrule
\end{tabular}
\caption{Descriptive statistics from the Claude-Qwen audit JSON files. Sup. = supported, Part. = partially supported, Not = not supported, Insuf. = insufficient evidence, Conf. = mean confidence, Retr. = mean retrieval strength, and Spec. = mean evidence specificity.}
\label{tab:audit-json-summary}
\end{table}

The audit-file labels show that the model did not simply mark most comments as supported. Of the 80 records, 13 were labelled \texttt{supported}, 32 were labelled \texttt{partially\_supported}, 33 were labelled \texttt{not\_supported}, and 2 were labelled \texttt{insufficient\_evidence}. This distribution is useful for the editorial workspace because it creates a range of cases for inspection: clear support, partial support, absence of support, and missing evidence.

\begin{table}[t]
\centering
\small
\begin{tabular}{lrrrrr}
\toprule
\textbf{Model label} & \textbf{Items} & \textbf{Mean conf.} & \textbf{Mean retr.} & \textbf{Mean spec.} & \textbf{Mean coverage} \\
\midrule
\texttt{supported} & 13 & 0.837 & 0.724 & 0.925 & 0.528 \\
\texttt{partially\_supported} & 32 & 0.685 & 0.740 & 0.691 & 0.542 \\
\texttt{not\_supported} & 33 & 0.463 & 0.615 & 0.198 & 0.388 \\
\texttt{insufficient\_evidence} & 2 & 0.387 & 0.615 & 0.075 & 0.200 \\
\bottomrule
\end{tabular}
\caption{Mean confidence components by Claude-Qwen support label. The pattern shows that higher confidence is associated mainly with higher evidence specificity, while insufficient-evidence cases have very low evidence specificity and low section coverage.}
\label{tab:audit-label-components}
\end{table}

These descriptive statistics support the design of the confidence score. The strongest separation between labels appears in evidence specificity: \texttt{supported} records have a mean evidence specificity of 0.925, while \texttt{not\_supported} records have a mean of 0.198 and \texttt{insufficient\_evidence} records have a mean of 0.075. This is consistent with the weighting choice in the prototype, where evidence specificity receives the largest weight. Retrieval strength is also informative, but it is less discriminating than evidence specificity. For example, \texttt{partially\_supported} records have the highest mean retrieval strength, but lower evidence specificity than \texttt{supported} records. This suggests that retrieving semantically close passages is not enough; the retrieved passages must also be specific to the reviewer comment.

At the paper level, Paper 2 has the highest mean confidence and mean evidence specificity. Paper 3 has the lowest mean confidence and the largest number of \texttt{not\_supported} labels. This is consistent with the qualitative observation that short, general, or subjective reviewer comments are harder to verify through retrieved manuscript evidence. The audit-file statistics therefore reinforce the main editorial point of the paper: Evidence-RAG is most useful when the reviewer comment can be connected to explicit, retrievable evidence, and it should be treated more cautiously when comments are broad, subjective, or outside the manuscript text.

\section{Limitations}
The current system is a research prototype. It should not be used as an autonomous peer-review agent. Several limitations remain. The confidence score is heuristic. Section detection and chunking can fail on unusual PDF structures. Retrieved evidence can be incomplete. Finally, the workflow has been tested on a limited corpus, so stronger general claims require further evaluation.

\section{Ethical and Editorial Position}
The workspace follows a human-in-the-loop editorial principle: AI may help organise evidence, but it must not replace scholarly judgement. The model output is treated as an object for inspection, not as a decision. The paper is also treated as protected editorial material. The workflow should minimise what is exposed to any model endpoint and should prefer evidence-bounded prompts over full-manuscript prompting. This distinction is essential in peer review, where fairness, accountability, and disciplinary expertise cannot be reduced to one automated score. Note that all sensitive information, including title, content and author names, has been masked.

\section{Conclusion}
This preliminary paper presents an Evidence-RAG workflow for making AI-assisted peer-review support more transparent and auditable. By connecting reviewer comments to retrieved manuscript evidence, model labels, confidence components, and editor decisions, the workspace aims to support fairer and more reproducible editorial practice. The next version will include broader experiments, refined evaluation, and a fuller discussion of how this approach can be integrated into the editorial workflow of the \emph{Journal of Digital History}.

\section*{Conference Note}
This preliminary work is connected to the accepted presentation ``Towards an Interactive `Evidence-RAG' Peer-Review Workspace for the Journal of Digital History'' at \emph{AI through History, History through AI}, hosted by C\textsuperscript{2}DH, University of Luxembourg, 15--16 June 2026.

\bibliographystyle{plainnat}
\bibliography{references}

\appendix

\section{Code and Repository}

The project code and supporting materials are available in the project repository:

\begin{center}
\url{https://github.com/C2DH/journal-of-digital-history-evidence-rag}
\end{center}

The repository contains the workflow scripts and notebooks used for paper conversion, review processing, vector database construction, Evidence-RAG auditing, and the editor-facing evaluation interface. The repository may be updated as the full version of this paper is completed.

\end{document}